\documentclass[aps,prl,twocolumn,10pt,showpacs,superscriptaddress,amsmath,amssymb]{revtex4-1}

\usepackage{graphicx}

\begin{document}

\title{Deterministic noiseless amplification of coherent states}
\author{Meng-Jun Hu}
\author{Yong-Sheng Zhang}\email{yshzhang@ustc.edu.cn}
\affiliation{Laboratory of Quantum Information, University of Science and Technology of China, Hefei, 230026, China}
\affiliation{Synergetic Innovation Center of Quantum Information and Quantum Physics, University of Science and Technology of China, Hefei, 230026, China}

\date{\today}

\begin{abstract}
A universal deterministic noiseless quantum amplifier has been shown to be impossible. However, probabilistic noiseless amplification of a certain set of states is physically permissible. Regarding quantum state amplification as quantum state transformation, we show that deterministic noiseless amplification of coherent states chosen from a proper set is possible. The relation between input coherent states and gain of amplification for deterministic noiseless amplification is thus derived. Besides, the potential applications of amplification of coherent states in quantum key distribution (QKD), noisy channel and non-ideal detection are also discussed.
\end{abstract}


\maketitle

Quantum amplification plays an essential role in quantum measurement and quantum metrology \cite{1,2}. In order to measure a weak signal, improving the sensitivity of the detector or amplifying the signal are two basic ways. However, constrained by physical laws, it may be very difficult for detectors to measure a weak enough signal especially for a quantum signal. What we usually do is amplifying the signal first and then measure it with proper detector. Unfortunately, the noise accompanying signal is also amplified during the process of signal amplification. What's worse is that the added noise may be brought in such that the signal-to-noise ratio (SNR) after amplification is reduced. For a linear phase-insensitive quantum amplifier, it has been shown that there is at least $(g^2-1/2)\hbar\omega$ total noise (including intrinsic noise and added noise) power per unit bandwidth out of its output-port, where $g^2$ is the power gain \cite{3,4}. The noiseless amplification (without introduce added noise) which is SNR-preserving seems unlikely for a universal linear phase-insensitive quantum amplifier.

In fact, due to the constraint of quantum commutation condition, a universal linear phase-insensitive quantum amplifier which can amplify any coherent states determinately and noiselessly is impossible \cite{5}. However, as the no-cloning theorem \cite{6,7,8,9} does not rule out the possibility of probabilistic cloning the state which is randomly chosen from a linear-independent set of states \cite{10}, the non-existence of a universal deterministic noiseless quantum amplifier does not mean the non-existence of a specific quantum amplifier which can noiselessly amplify a certain input set of states. The noiseless amplification of quantum states essentially is a problem of quantum states transformation. Both deterministic and probabilistic quantum states transformation have already been discussed in detail  \cite{11,12,13}. Using the language of quantum states transformation, the quantum states clone, the unambiguous discrimination of states and the quantum states amplification can be demonstrated in a uniform framework \cite{14,15,16}. Recently, there are some experimental reports on realization of noiseless amplification of quantum light states \cite{17,18,19}. All these experimental schemes are probabilistic and can only attain unit fidelity asymptotically. The truly probabilistic noiseless amplification of coherent states in a certain set is thus discussed based on quantum states transformation \cite{16}.

A natural question arises: Whether or not there exists a specific quantum amplifier which can determinately and noiselessly amplify a coherent state randomly chosen from a definite set of coherent states? In this paper, we will show that it is indeed possible if we regard quantum states amplification as quantum states transformation.

We note here that there is another kind of quantum amplifier which is based on the unambiguous identification of input states and the preparation of desired amplified input state \cite{20,21}. This kind of quantum amplifier is usually called classic-like quantum amplifier analogous to classical amplifier working through measurement and preparation. Limited by success probability of identification of input states(except orthogonal input states), the classic-like quantum amplifier can only probabilistic amplify the input states though the gain of amplification can be arbitrarily high.

{\it Quantum transformation of sets of pure states.}
In quantum operation theory, any physically permissible transformation of the state of a quantum system can be represented by a completely positive (CP), linear, trace non-increasing map: $\$:\rho\rightarrow\$(\rho)$. If any such map exists, the transformation is realizable in principle \cite{22}. There are two mostly used equivalent ways to illustrate such a transformation. The so-called first representation theorem \cite{23} states that the CP, linear, trace non-increasing map $\$$ can be represented as operator-sum form $\$(\rho)=\sum_{k}A_{k}^{\dagger}\rho A_{k}$, where $A_{k}$ is the Kraus operator which satisfy $\sum_{k}A_{k}^{\dagger}A_{k}\leq I$, $I$ is the identity operator. For deterministic transformation $\sum_{k}A_{k}^{\dagger}A_{k}=I$, while for probabilistic transformation $\sum_{k}A_{k}^{\dagger}A_{k}<I$. This process has another description. It says that the transformation can be implemented by adding an ancillary system to the quantum system and then a unitary transformation is applied to the composite system. Mathematically, we have $\$(\rho)=tr_{E^{'}}[U\rho\bigotimes\rho_{E} U^{\dagger}I\bigotimes P_{E^{'}}]$, where $\rho_{E}$ is the initial state of ancillary system, $P_{E^{'}}$ is a projector in transformed ancillary Hilbert space \cite{24}. We begin our formal discussion by first review the well-known quantum transform theorem of sets of pure states.

{\textbf{Lemma 1} \cite{13}\textbf{:}} Suppose there is a set of $N$ pure states $A=\lbrace |\psi_{i}\rangle\rbrace$ which is linear-independent and another set of $N$ pure states $B=\lbrace |\phi_{i}\rangle\rbrace$. A probabilistic transformation $T:A=\lbrace |\psi_{i}\rangle\rbrace\xrightarrow{\lbrace p_{i}\rbrace} B=\lbrace |\phi_{i}\rangle\rbrace$ that transform state $|\psi_{i}\rangle$ in set $A$ to the corresponding state $|\phi_{i}\rangle$ in set $B$ with probability $p_{i}$ exists if and only if there exists an $N\times N$ matrix $\Pi$ which satisfy the three conditions: $1:\Pi\ge 0$; $2:diag(\Pi)=\vec{p}=(p_{1},p_{2},...,p_{N})$; $3:G_{A}-\Pi\circ G_{B}\ge 0$. Here $G_{A}$ and $G_{B}$ are Gram matrix of set $A$ and $B$ respectively and $\circ$ denotes Hadamard matrix product.

{\textbf{Proof:}} If such a transformation exists, there must exist complex coefficients $c_{ki}$ such that
\begin{equation}
A_{ks}|\psi_{i}\rangle=c_{ki}|\phi_{i}\rangle,
\end{equation}
where $A_{ks}(k=1,2,...,M)$ are the Kraus operators for successful transformation.
Consider these coefficients as the elements of a $M\times N$ matrix $C=[c_{ki}]$. Now we can introduce matrix $\Pi$ defined by: $\Pi=C^{\dagger}C$. It can be easily shown that matrix $\Pi$ satisfies all three conditions \cite{13}.

Suppose there is a matrix $\Pi$ which satisfies all three conditions. Positivity of the matrix enable us to factorize $\Pi$ as $C^{\dagger}C$ and then the transformation operators can be constructed as
\begin{equation}
A_{ks}=\sum_{i}\frac{c_{ki}}{\langle\tilde{\psi_{i}}|\psi_{i}\rangle}|\phi_{i}\rangle\langle\tilde{\psi_{i}}|,
\end{equation}
where $\langle\tilde{\psi_{i}}|\psi_{j}\rangle=\gamma_{i}\delta_{ij}$, $\gamma_{i}\neq 0$ is a constant. State $|\tilde{\psi_{i}}\rangle$ is orthogonal to any state in set $A$ except for state $|\psi_{i}\rangle$. 

Physically, the above transformation can be implemented by a specific unitary transformation operating on a composite system consists of quantum system and ancillary system \cite{16}.
\begin{equation}
U|\psi_{i}\rangle=\sqrt{p_{i}}|\phi_{i}\rangle|u_{i}\rangle|0\rangle+\sqrt{1-p_{i}}|Fail\rangle|v_{i}\rangle|1\rangle.
\end{equation}
Taking the inner product of Eq.(3) and its complex conjugate, we have:
\begin{equation}
\langle\psi_{j}|\psi_{i}\rangle=\sqrt{p_{i}p_{j}}\langle\phi_{j}|\phi_{i}\rangle\langle u_{j}|u_{i}\rangle+\sqrt{(1-p_{i})(1-p_{j})}\langle v_{j}|v_{i}\rangle,
\end{equation}
which can be recast as:
\begin{equation}
G_{A}=G_{B}\circ\Pi+K.
\end{equation}
From the positivity of Gram matrix $K$, it is easily to see that $G_{A}-G_{B}\circ\Pi\ge 0$ and the Gram matrix $\Pi$ defined as $\Pi=\sqrt{p_{i}p_{j}}\langle u_{j}|u_{i}\rangle$ obviously satisfies the conditions 1 and 2.

Consider the deterministic transformation which means $p_{i}=1$ for all states so that the second term in the right hand side of Eq.(4) vanish. Eq.(5) becomes $G_{A}=G_{B}\circ\Pi$ with $\Pi=\langle u_{j}|u_{i}\rangle$. For any two input states in the set $A$, we have the following equality:
\begin{equation}
\langle\psi_{j}|\psi_{i}\rangle=\langle\phi_{j}|\phi_{i}\rangle\langle u_{j}|u_{i}\rangle.
\end{equation}
Eq.(6) implies that the overlap between two input states is no more than the overlap between two corresponding output states after deterministic transformation, that is $|\langle\psi_{j}|\psi_{i}|\leq |\langle\phi_{j}|\phi_{i}\rangle|$. From the point view of information, the deterministic transformation does not increase the distinguishability of input states. In fact, any physical process of states does not increase our knowledge of states.

\begin{figure}[tbp]
\centering
\includegraphics[scale=0.2]{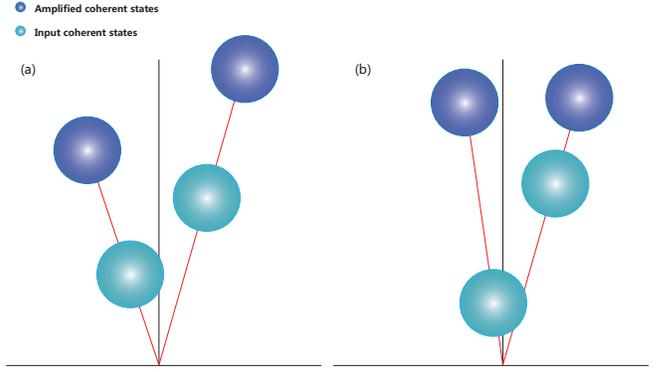}
\caption{(colour online)  {\it Illustration of Wigner function contours for probabilistic and deterministic noiseless amplification of coherent states.} (a) The distance between two amplified coherent states is longer than the distance between two  input coherent states, the noiseless amplification of coherent states can only be probabilistic. (b) The distance between two amplified coherent states is shorter than or equal to the distance between two input coherent states, the noiseless amplification of coherent states can be deterministic as long as the gain of amplification of quantum amplifier is state-dependent.   }
\end{figure}

{\it Deterministic noiseless amplification of coherent states.}
We now focus on the case of coherent states. The above discussion can be directly applied to the set of coherent states. In the space of coherent states, the distinguishability of two coherent states can be measured by the distance of two coherent states, since that $|\left\langle \alpha_{1}|\alpha_{2}\right\rangle |^{2}=exp(-|\alpha_{1}-\alpha_{2}|^{2})$. We can define the distance of two coherent states $|\alpha_{1}\rangle$ and $|\alpha_{2}\rangle$ as
\begin{equation}
D(\alpha_{1},\alpha_{2})=|\alpha_{1}-\alpha_{2}|^{2}.
\end{equation}
The distinguishability of any two coherent states is thus proportional to their distance. If there exists a deterministic transformation which transform the set of coherent states $A=\lbrace |\alpha_{i}\rangle\rbrace$ to another set of coherent states $B=\lbrace |\beta_{i}\rangle\rbrace$, we must have $D(\alpha_{i},\alpha_{j})\ge D(\beta_{i},\beta_{j})$. However, it is not hold for a deterministic noiseless quantum amplifier of coherent states with gain of amplification $g>1$. It is obviously that $D(g\alpha_{i},g\alpha_{j})=g^{2}|\alpha_{i}-\alpha_{j}|^{2}>|\alpha_{i}-\alpha_{j}|^{2}=D(\alpha_{i},\alpha_{j})$. For the quantum amplifier of coherent states with fixed gain of amplification $g>1$, deterministic noiseless amplification is thus impossible. From the point view of quantum states transformation, however, a fixed gain of amplification is not necessary. As shown in Fig.1, a deterministic noiseless quantum amplifier of coherent states can exist as long as the gain of amplification is state-dependent. For the simplicity of discussion, we just consider the case of only two coherent states contained in the input set in the following.

{\textbf{Theorem 1:}} 
Suppose there are two sets of two coherent states $A=\lbrace |\alpha_{1}e^{i\theta_{1}}\rangle,|\alpha_{2}e^{i\theta_{2}}\rangle\rbrace$, $B=\lbrace |g_{1}\alpha_{1}e^{i\theta_{1}}\rangle,|g_{2}\alpha_{2}e^{i\theta_{2}}\rangle\rbrace$, here we explicitly show the amplitude and phase of coherent state. The deterministic noiseless quantum amplifier which amplify the coherent state randomly chosen from set $A$ to the corresponding coherent state in the set $B$ exists if and only if the input coherent states and the gain of amplification satisfy the condition: $\cos\eta\ge\sqrt{(g_{1}^{2}-1)(g_{2}^{2}-1)}/(g_{1}g_{2}-1)$ with $\eta=|\theta_{1}-\theta_{2}|$ is the relative phase between two coherent states.

{\textbf{Proof:}}
If the deterministic noiseless amplification exists then we must have
\begin{equation}
D(g_{1}\alpha_{1}e^{i\theta_{1}},g_{2}\alpha_{2}e^{i\theta_{2}})\le D(\alpha_{1}e^{i\theta_{1}},\alpha_{2}e^{i\theta_{2}}).
\end{equation}
According to the definition of the distance of two coherent states, we have
\begin{equation}
|\alpha_{1}e^{i\theta_{1}}-\alpha_{2}e^{i\theta_{2}}|^{2}\ge |g_{1}\alpha_{1}e^{i\theta_{1}}-g_{2}\alpha_{2}e^{i\theta_{2}}|^{2}.
\end{equation} 
The calculation of Eq.(9) gives $2\alpha_{1}\alpha_{2}(g_{1}g_{2}-1)\cos\eta\ge (g_{1}^{2}-1)\alpha_{1}^{2}+(g_{2}^{2}-1)\alpha_{2}^{2}\ge 2\alpha_{1}\alpha_{2}\sqrt{(g_{1}^{2}-1)(g_{2}^{2}-1)}$. Eliminating the same factor in both sides finally gives
\begin{equation}
\cos\eta\ge\frac{\sqrt{(g_{1}^{2}-1)(g_{2}^{2}-1)}}{g_{1}g_{2}-1}.
\end{equation}
The equality holds only when $\sqrt{g_{1}^{2}-1}\alpha_{1}=\sqrt{g_{2}^{2}-1}\alpha_{2}$.

If the input coherent states and gain of amplification satisfy Eq.(10), we can always construct Kraus operator of amplification as
\begin{equation}
A_{k}=\frac{c_{k1}}{\langle\tilde{\psi_{1}}|\alpha_{1}e^{i\theta_{1}}\rangle}|g_{1}\alpha_{1}e^{i\theta_{1}}\rangle\langle\tilde{\psi_{1}}|+\frac{c_{k2}}{\langle\tilde{\psi_{2}}|\alpha_{2}e^{i\theta_{2}}\rangle}|g_{2}\alpha_{2}e^{i\theta_{2}}\rangle\langle\tilde{\psi_{2}}|,
\end{equation}
where $\sum_{k}A_{k}^{\dagger}A_{k}=I$ and $\lbrace|\tilde{\psi_{1}}\rangle$, $|\tilde{\psi_{1}}\rangle\rbrace$ are defined as
\begin{equation}
|\tilde{\psi_{1}}\rangle=\frac{1}{\langle\alpha_{2}e^{i\theta_{2}}|\alpha_{1}e^{i\theta_{1}}\rangle}|\alpha_{1}e^{i\theta_{1}}\rangle-|\alpha_{2}e^{i\theta_{2}}\rangle,
\end{equation}
\begin{equation}
|\tilde{\psi_{2}}\rangle=\frac{1}{\langle\alpha_{1}e^{i\theta_{1}}|\alpha_{2}e^{i\theta_{2}}\rangle}|\alpha_{2}e^{i\theta_{2}}\rangle-|\alpha_{1}e^{i\theta_{1}}\rangle.
\end{equation}
Notice that
\begin{equation}
\langle\tilde{\psi_{s}}|\alpha_{t}e^{i\theta_{t}}\rangle=\frac{1-|\langle\alpha_{1}e^{i\theta_{1}}|\alpha_{2}e^{i\theta_{2}}\rangle|^{2}}{\langle\alpha_{t}e^{i\theta_{t}}|\alpha_{s}e^{i\theta_{s}}\rangle}\delta_{s,t}, 
\end{equation}
where $s,t=1,2$, $\delta_{s,t}=1$ for $s=t$ and $\delta_{s,t}=0$ for $s\neq t$.
It can be easily verified that
\begin{equation}
A_{k}|\alpha_{1}e^{i\theta_{1}}\rangle=c_{k1}|g_{1}\alpha_{1}e^{i\theta_{1}}\rangle,
\end{equation}
\begin{equation}
A_{k}|\alpha_{2}e^{i\theta_{2}}\rangle=c_{k2}|g_{2}\alpha_{2}e^{i\theta_{2}}\rangle.
\end{equation}
The Kraus operator $A_{k}$ we construct is indeed the expected operator of amplification.

The result above can be easily extended into the more general case in which the input set contains more than two coherent states. For that case, a deterministic noiseless quantum amplifier exists if and only if any two coherent states in the set satisfy the relation (10). There is a specific amplification after which all the amplified coherent states have the same amplitude. In two coherent states case, it means $g_{1}\alpha_{1}=g_{2}\alpha_{2}$ and the input states and gain of amplification have to satisfy more restricted relation. We thus have the following corollary.



{\textbf{Corollary 1:}}
Suppose there are two sets of two coherent states $A=\lbrace |\alpha_{1}e^{i\theta_{1}}\rangle,|\alpha_{2}e^{i\theta_{2}}\rangle\rbrace$, $B=\lbrace |g_{1}\alpha_{1}e^{i\theta_{1}}\rangle,|g_{2}\alpha_{2}e^{i\theta_{2}}\rangle\rbrace$ with $g_{1}\alpha_{1}=g_{2}\alpha_{2}$. The deterministic noiseless quantum amplifier which amplify the coherent state randomly chosen from set $A$ to the corresponding coherent state in the set $B$ exists if and only if the input coherent states and the gain of amplification satisfy the condition $\cos\eta\ge\frac{(2g_{1}^{2}-1)\alpha_{1}^{2}-\alpha_{2}^{2}}{2g_{1}^{2}\alpha_{1}^{2}-2\alpha_{1}\alpha_{2}}$.

{\textbf{Proof:}}
The proof is as same as in theorem 2. The only difference is that the input states and gain of amplification need satisfy more restricted relation due to the requirement of $g_{1}\alpha_{1}=g_{2}\alpha_{2}$. Substituting $g_{1}\alpha_{1}=g_{2}\alpha_{2}$ into the Eq.(9), we thus get
\begin{equation}
\cos\eta\ge\frac{(2g_{1}^{2}-1)\alpha_{1}^{2}-\alpha_{2}^{2}}{2g_{1}^{2}\alpha_{1}^{2}-2\alpha_{1}\alpha_{2}}.
\end{equation}

Among deterministic noiseless quantum amplifiers which amplify coherent states to the same final amplitude, the best deterministic noiseless quantum amplifier for a definite input set of coherent states can be defined as $g_{1}$ has the maximum value. The maximum value of $g_{1}$ can be easily calculated from Eq.(17)
\begin{equation}
g_{1max}=\sqrt{\frac{\alpha_{1}^{2}+\alpha_{2}^{2}-2\alpha_{1}\alpha_{2}\cos\eta}{2\alpha_{1}^{2}(1-\cos\eta)}}.
\end{equation}

{\it Discussion and summary.}
According to the theory of quantum states transformation, we have shown that the deterministic noiseless amplification of coherent states is theoretically possible. In general, the distinguishability of two coherent states after deterministic noiseless amplification is reduced. However, in the task of continuous variable quantum key distribution (QKD) using coherent light pulses \cite{25}, the reduced distinguishability of amplified coherent states is not a prominent defect. Conversely, the amplification of coherent light pulses can not only increase the transmission distance but also beneficial to the improvement of signal transmission rate.

Though the distinguishability of coherent states does not increase after deterministic noiseless amplification, the situation may be different in a noisy channel. Consider two coherent states $|\alpha_{1}e^{i\theta_{1}}\rangle$ and $|\alpha_{2}e^{i\theta_{2}}\rangle$ which satisfy the Eq.(10) such that they can be deterministic noiseless amplified to corresponding coherent states $|g_{1}\alpha_{1}e^{i\theta_{1}}\rangle$ and $|g_{2}\alpha_{2}e^{i\theta_{2}}\rangle$ respectively. When the amplified coherent states are sent through a noisy channel then, the distinguishability of two amplified states through the noisy channel may be  larger than the two coherent states through the same noisy channel without amplification. To see this more explicitly, using a superoperator $V(t)$ to describe the noisy channel such that the state evolution of quantum system in noisy channel is $\rho(t)=V(t)[\rho(0)]$. For two coherent states $\rho_{1}(0)=|\alpha_{1}e^{i\theta_{1}}\rangle\langle\alpha_{1}e^{i\theta_{1}}|$ and $\rho_{2}(0)=|\alpha_{2}e^{i\theta_{2}}\rangle\langle\alpha_{2}e^{i\theta_{2}}|$, the distance of two states in noisy channel is in general decreased monotonously $D(\rho_{1}(t),\rho_{2}(t))\leq D(\rho_{1}(0),\rho_{2}(0)$. We thus can define the decay rate of distance as $\sigma(\rho_{1}(t),\rho_{2}(t))=\frac{d}{dt}D(\rho_{1}(t),\rho_{2}(t))$. Obviously, $\sigma(\rho_{1}(t),\rho_{2}(t))\leq 0$ means the distinguishability of two states decrease with time in noisy channel. Similarly, we can obtain the decay rate of two amplified coherent states in same noisy channel $\sigma(\rho_{1}^{g_{1}}(t),\rho_{2}^{g_{2}}(t))=\frac{d}{dt}D(\rho_{1}^{g_{1}}(t),\rho_{2}^{g_{2}}(t))$, where $\rho_{1}^{g_{1}}(t)=|g_{1}\alpha_{1}e^{i\theta_{1}}\rangle\langle g_{1}\alpha_{1}e^{i\theta_{1}}|$ and $\rho_{2}^{g_{2}}(t)=|g_{2}\alpha_{2}e^{i\theta_{2}}\rangle\langle g_{2}\alpha_{2}e^{i\theta_{2}}|$ are amplified coherent states. In general, the decay rate of distance depend on the initial states which means the decay rate may be different for coherent states and amplified coherent states. It thus possible that the distinguishability of two amplified coherent states through the noisy channel may be larger than the two coherent states through the same noisy channel without amplification. Besides, the detectors we use to distinguish coherent states are not truly ideal in practice. The unavoidable dark noise will cause dark counting in the detector which lower our precision of distinguish coherent states. For a non-ideal detector, the amplified coherent states may be more distinguishable than the coherent states without amplification. As an exactly solvable noisy model have not been found  for deeper investigation, we leave it as an open question whether or not amplified coherent states perform better in noisy environment.

In conclusion, we have shown that the deterministic noiseless amplification of coherent states is physically possible if we regard the quantum states amplification as quantum states transformation. Our results are based on two facts: the process of deterministic noiseless amplification does not increase the distinguishability of any two amplified states and the gain of amplification can be state-dependent. The relation between input coherent states and gain of amplification for deterministic noiseless amplification is thus derived. We also discussed the potential applications in QKD, noisy channel and non-ideal detection. The result we obtain may be useful in quantum communication using coherent states of light as information carrier.

We thank X. F. Zhou and G. Y. Xiang for helpful discussions. This work is supported by National Natural Science Foundation of China (No.61275122), National Fundamental Research Program of China and Strategic Priority Research Program (B) of CAS (No.XDB01030200).

\end{document}